\documentclass[11pt]{article}
\def \d {{\rm d}}

\begin{document}

\title{The stability of Killing--Cauchy horizons \\
in colliding plane wave space-times}

\author{ J. B. Griffiths\thanks{E--mail: {\tt J.B.Griffiths@Lboro.ac.uk}} \\ \\ 
Department of Mathematical Sciences, Loughborough University, \\ 
Loughborough, Leics. LE11 3TU, U.K. \\ \\}

\date{\today}
\maketitle

\begin{abstract}
\noindent
It is confirmed rigorously that the Killing--Cauchy horizons, which sometimes
occur in space-times representing the collision and subsequent interaction of
plane gravitational waves in a Minkowski background, are {\em unstable} with
respect to {\em bounded} perturbations of the initial waves, at least for the
case in which the initial waves have constant aligned polarizations. 
\end{abstract}

\section{Introduction}
Many classes of explicit exact solutions are known which model the collision and
subsequent interaction between shock-fronted plane gravitational waves which
propagate and collide in a Minkowski background (for reviews see \cite{Griff91}
or \cite{SKMHH03}). In all these solutions, some kind of singularity always
appears in the interaction region. This is generally a spacelike curvature
singularity (like the time-reverse of an initial cosmological singularity).
However large classes of solutions with an infinite number of parameters exist in
which the scalar polynomial curvature singularity is replaced by a horizon. In
these cases, the space-time can be extended through the horizon, but the
extension is not unique. It has become widely believed that such Killing--Cauchy
horizons are unstable with respect to small changes in the initial data. For
example, Yurtsever \cite{Yurts87} has shown that they are unstable with respect
to the addition of some perturbative linear field which preserves the $G_2$
symmetry. However, this result does not answer the question of whether they are
unstable with respect to variations in the approaching purely gravitational waves
when the vacuum field equations are satisfied exactly. It is the purpose of the
present paper to investigate this question in detail.

It must first be pointed out that, in the vast amount of work that was
undertaken on this topic in the 1970s and 1980s, and in all the exact solutions
that were then produced, the approach was adopted of first solving the field
equations in the interaction region. Once a family of such solutions had been
obtained, the free parameters were constrained to satisfy the junction conditions
that have to be imposed in order to extend the solution to the prior regions.
Thus, the approaching waves which physically give rise to the solutions were only
determined once the solution had been obtained. Within this context, it was argued
that those solutions which contained horizons were unstable with respect to
perturbations of the initial data. However, it was not then appreciated that the
perturbations which transform a Killing--Cauchy horizon to a scalar polynomial
curvature singularity also normally introduce singularities in the initial waves
prior to their collision. The question therefore still needs to be addressed as to
whether or not the horizons are stable with respect to regular (bounded)
perturbations of physically acceptable initial waves.

It is only very recently that techniques have been developed by which colliding
plane wave solutions can be explicitly constructed from their characteristic
initial data representing the approaching waves \cite{HauErn89a}--\cite{AleGri04}.
It is therefore only now that tools are available to reconsider this question. In
this paper, only the linear vacuum case in which two approaching gravitational
waves have constant aligned polarizations will be considered. This situation is
much easier to analyse, but it already demonstrates the essential features of the
physical situation. The basic result, not surprisingly, is that the
Killing--Cauchy horizons that sometimes appear in colliding plane wave
space-times are unstable with respect to bounded perturbations of the initial
waves that generate these solutions.

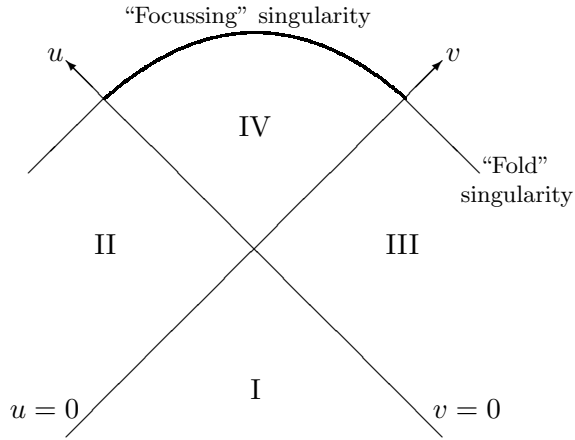
\begin{figure}[hpt]      
\begin{center}
\setlength{\unitlength}{0.25mm}
\begin{picture}(200,250)(-100,-100)
\put(-100,-100){\vector(1,1){200}}
\put(100,-100){\vector(-1,1){200}}
\put(-80,80){\line(-1,-1){40}}
\put(80,80){\line(1,-1){40}}
\put(-110,100){$u$}
\put(102,100){$v$}
\put(-130,-90){$u=0$}
\put(95,-90){$v=0$}
\put(-2,-80){I}
\put(-85,-3){II}
\put(70,-3){III}
\put(-8,60){IV}
\put(-70,120){\fontsize{9}{14}\selectfont ``Focussing'' singularity}
\put(120,40){\fontsize{9}{14}\selectfont ``Fold''}
\put(110,25){\fontsize{9}{14}\selectfont singularity}
\thicklines
\qbezier(-80,80)(0,150)(80,80)
\end{picture}
\vskip-0.6cm
\end{center}

\caption{\small The structure of colliding plane wave space-times. Region~I is the
background Minkowski space, regions II and~III contain the approaching plane
waves, and region~IV represents the interaction region following the collision.
The singularity structure following the collision is known from the study of
families of exact solutions and is described in \cite{Griff91}.}
\end{figure}

\section{Initial data}

Colliding plane wave space-times are naturally divided into four regions as
indicated in figure~1. It is found to be convenient to use two future-pointing
null coordinates $u$ and $v$ throughout the space-time. The four distinct regions
can then be identified as those that are separated by two null hypersurfaces
taken as $u=0$ and $v=0$ which represent the wavefronts of the approaching
waves.

The background region~I ($u<0$, $v<0$) is a flat vacuum represented by the line
element 
 \begin{equation}
 \d s^2=2\,\d u\,\d v-\d x^2-\d y^2. 
 \label{Minkowski}
 \end{equation}

Region~II ($u\ge0$, $v<0$) contains one of the approaching plane waves. If this
wave has constant (linear) polarization, it can be described by a metric in the
Brinkmann form 
 \begin{equation}
 \d s^2 =2\,\d u\,\d r -\d X^2 -\d Y^2 
+h_{\scriptscriptstyle+}(u)(X^2-Y^2)\,\d u^2, 
 \label{Brinkmann}
 \end{equation} 
 where $h_{\scriptscriptstyle+}(u)=\Psi_{4\scriptscriptstyle+}(u)$, which is the
only non-zero component of the Weyl tensor (relative to an appropriate tetrad). 
This arbitrary function explicitly represents the profile of the plane
gravitational wave as a function of the retarded time~$u$. (The subscripts~$+$
and~$-$ are used here and below to denote functions that are defined in
regions~II and~III respectively.)

The form of the metric (\ref{Brinkmann}) may be taken to include region~I with
$h_{\scriptscriptstyle+}=0$ for $u<0$. However, to analyse the collision of plane
waves, it is appropriate to initially transform the line element
(\ref{Brinkmann}) to the form 
 \begin{equation}
 \d s^2=2\,\d u\,\d v  -e^{2\psi_+}\,\d x^2
-\alpha_{\scriptscriptstyle+}^2\,e^{-2\psi_+}\,\d y^2. 
 \label{metric2}
 \end{equation}
 As explained in \cite{AleGri04}, the coordinate $u$ is an affine parameter on
the null geodesics \ $v=$~const. \ $x=$~const. \ $y=$~const. \ which cross the
wavefront from the background region. Also, the function
$\psi_{\scriptscriptstyle+}(u)$ must satisfy the linear scattering equation 
 \begin{equation}
 (e^{\psi_+})_{,uu}
+h_{\scriptscriptstyle+}(u)\,e^{\psi_+}=0. 
 \label{P}
 \end{equation} 
 In order to join (\ref{metric2}) smoothly with (\ref{Minkowski}), it is
necessary that \ $\psi_{\scriptscriptstyle+}(0)=0$ \ and that
$\psi_{{\scriptscriptstyle+},u}(0)$ must vanish except for any non-zero component
which arises from a possible impulsive component on the wavefront. With these
conditions, $\psi_{\scriptscriptstyle+}(u)$ is determined uniquely from the
initial wave profile $h_{\scriptscriptstyle+}(u)$. The function
$\psi_{\scriptscriptstyle+}(u)$ can therefore be taken to represent the initial
data to be prescribed on the characteristic $v=0$.

With $\psi_{\scriptscriptstyle+}(u)$ determined, the initial gravitational wave
is given by the Weyl tensor component 
 \begin{equation}
 \Psi_{4\scriptscriptstyle+}(u)= 
-\psi_{{\scriptscriptstyle+},u}\delta(u)
-\left[ \psi_{{\scriptscriptstyle+},uu} 
+(\psi_{{\scriptscriptstyle+},u})^2 \right]\Theta(u) . 
 \label{h(u)}
 \end{equation}
 The other metric function $\alpha_{\scriptscriptstyle+}(u)$ in (\ref{metric2})
is then determined for region~II from the linear equation 
 \begin{equation} 
 \alpha_{{\scriptscriptstyle+},uu}
 -2\,\psi_{{\scriptscriptstyle+},u}\, \alpha_{{\scriptscriptstyle+},u} 
 +2(\psi_{{\scriptscriptstyle+},u})^2\, \alpha_{\scriptscriptstyle+}=0, 
 \label{alphaplus}
 \end{equation} 
 whose solution is uniquely specified by the initial data \
$\alpha_{\scriptscriptstyle+}(0)=1$, \ and \
$\alpha_{{\scriptscriptstyle+},u}(0)=0$. \ It can be seen that
$\alpha_{\scriptscriptstyle+}(u)$ must be a monotonically decreasing function in
this region.

Region~III ($u<0$, $v\ge0$) contains the other plane wave which approaches from
the opposite direction. If this wave also has constant polarization which is
aligned with that in region~II, the metric can simply be taken as having the same
form as that in that region, but with the roles of $u$ and $v$ reversed. Thus, the
initial data that is specified in this region will be taken as the function
$\psi_{\scriptscriptstyle-}(v)$. The non-zero component of the Weyl tensor
$\Psi_{0\scriptscriptstyle-}(v)$ representing the second initial gravitational
wave is given by an equation equivalent to (\ref{h(u)}), and the metric function
$\alpha_{\scriptscriptstyle-}(v)$ is given by the equivalent of (\ref{alphaplus})
and the corresponding initial data.

Region~IV ($u\ge0$, $v\ge0$) represents the region in which the waves interact. 
In the vacuum case for the collision of plane gravitational waves with constant
aligned polarization, this region can always be described by the line element 
 \begin{equation}
 \d s^2=2\,f\,\d u\,\d v -e^{2\psi}\,\d x^2
-\alpha^2\,e^{-2\psi}\,\d y^2, 
 \label{metric}
 \end{equation}
 where $\psi(u,v)$, $\alpha(u,v)$ and $f(u,v)$ are now functions of both null
coordinates. If characteristic initial data is taken as described above, these
functions must satisfy the initial data:  
 \begin{equation}
 \begin{array}{c}
 \psi(u,0)=\psi_{\scriptscriptstyle+}(u) \\[1pt]
 \psi(0,v)=\psi_{\scriptscriptstyle-}(v)  
 \end{array} \qquad\qquad
\begin{array}{c}
 \alpha(u,0)=\alpha_{\scriptscriptstyle+}(u) \\[1pt]
 \alpha(0,v)=\alpha_{\scriptscriptstyle-}(v)  
 \end{array} \qquad\qquad
\begin{array}{c}
 f(u,0)=1  \\[1pt]
 f(0,v)=1 
 \end{array} 
 \label{junction} 
 \end{equation}
 with $\psi(0,0)=0$, $\alpha(0,0)=1$ and $f(0,0)=1$. \ The necessary (vacuum)
field equations will be considered in the following section.

\section{The field equations}

One of the vacuum field equations for the metric (\ref{metric}) is 
 $$ \alpha_{,uv}=0, $$ 
 whose solution, satisfying the necessary initial conditions, is given by 
 \begin{equation}
 \alpha(u,v)=\alpha_{\scriptscriptstyle+}(u) 
+\alpha_{\scriptscriptstyle+}(u) -1. 
 \label{alpha}
 \end{equation} 
 It is convenient to introduce the functions $\xi(u)$ and $\eta(v)$ such that 
$2\alpha=\xi(u)-\eta(v)$, where 
 $$ \xi(u)=\left\{ \begin{array}{l}
 1 \hskip4.5pc \hbox{in regions I and III} \\
 2\alpha_{\scriptscriptstyle+}(u)-1 \ \ \hbox{in regions II and IV} 
\end{array} \right. , \qquad
 \eta(v)=\left\{ \begin{array}{l}
 -1 \hskip3.7pc \hbox{in regions I and II} \\
 1-2\alpha_{\scriptscriptstyle-}(v) \ \ \hbox{in regions III and IV} 
\end{array} \right.  $$ 
 Then, since $\alpha_{\scriptscriptstyle+}(u)$ and
$\alpha_{\scriptscriptstyle-}(v)$ are monotonically decreasing, $\xi(u)$ must be
a monotonically decreasing function and $\eta(v)$ monotonically increasing in
the interaction region.

The other main vacuum field equation is 
 \begin{equation}
2\,\alpha\,\psi_{,uv} +\alpha_{,u}\,\psi_{,v} +\alpha_{,v}\,\psi_{,u}=0,
 \label{main}
 \end{equation}
 where $\alpha$ is given by (\ref{alpha}) and the initial data by
(\ref{junction}). For any solution of (\ref{main}) satisfying the initial data,
the additional metric function $f$ can be obtained by integrating the remaining
field equations which are 
 \begin{equation}
 {f_{,v}\over f} ={\alpha_{,vv}\over\alpha_{,v}}
+{2\,\alpha\,{\psi_{,v}}^2\over\alpha_{,v}} -2\,\psi_{,v}, \qquad\qquad 
 {f_{,u}\over f} ={\alpha_{,uu}\over\alpha_{,u}}
+{2\,\alpha\,{\psi_{,u}}^2\over\alpha_{,u}} -2\,\psi_{,u}. 
 \label{auxiliary}
 \end{equation} 
 The non-zero components of the Weyl tensor in the interaction region are then 
 \begin{equation}
 \begin{array}{l}
 \Psi_0= {\displaystyle -\psi_{,v}\delta(v) -{\psi_{,vv} +{\psi_{,v}}^2\over f} 
+{2\,\psi_{,v}\,f_{,v}\over f^2} } \\[8pt]
 \Psi_2={\displaystyle {1\over2\alpha f}\,\big( 2\alpha\,\psi_{,v}\,\psi_{,u}
-\alpha_{,u}\,\psi_{,v} -\alpha_{,v}\,\psi_{,u} \big)} \\[8pt]
 \Psi_4= {\displaystyle -\psi_{,u}\delta(u) -{\psi_{,uu} +{\psi_{,u}}^2\over f}
+{2\,\psi_{,u}\,f_{,u}\over f^2} }
 \end{array}
 \label{WeylCompts1}
 \end{equation} 
 It may be observed that a solution can be determined in region~IV up to the
spacelike hypersurface on which \ $\alpha={1\over2}(\xi-\eta)=0$, \ on which some
kind of singularity will occur. This is referred to as a ``focussing''
singularity in figure~1. Its character will be discussed below.

In view of the properties described above, it is always possible to adopt $\xi$
and $\eta$ as coordinates throughout the interaction region, in which \ $\xi<1$, \
$\eta>-1$ \ and \ $\xi-\eta>0$. \ In terms of these coordinates, the main field
equation (\ref{main}) becomes
 \begin{equation}
(\xi-\eta)\,\psi_{,\xi\eta} -{\textstyle{1\over2}}\,\psi_{,\xi}
+{\textstyle{1\over2}}\,\psi_{,\eta}=0,
 \label{EPD}
 \end{equation}
 which is an Euler--Poisson--Darboux equation with non-integer coefficients,
whose solution $\psi(\xi,\eta)$ must satisfy the initial data 
 \begin{equation}
 \psi(\xi,-1)=\psi_{\scriptscriptstyle+}(\xi), 
\qquad \psi(1,\eta)=\psi_{\scriptscriptstyle-}(\eta), 
 \label{Vconditions}
 \end{equation}
 with \ $\psi(1,-1)=0$. \ This is sufficient to uniquely determine the solution
of the colliding plane wave problem in the interaction region. However, as
explained in~\cite{AleGri04}, to satisfy the initial data, $\psi(\xi,\eta)$ must
be nonanalytic on each wavefront where it behaves like $\sqrt{1-\xi}$ or
$\sqrt{1+\eta}$ respectively. Also, in terms of these coordinates, the non-zero
Weyl tensor components are  
 \begin{equation}
 \begin{array}{l}
 {\displaystyle \Psi_0 =-\eta_{,v}\,\psi_{,\eta}\delta(v)
-{{\eta_{,v}^2}\over f}\left(\psi_{,\eta\eta}
+2(\xi-\eta)\,{\psi_{,\eta}}^3 +3{\psi_{,\eta}}^2\right)} \\[10pt]
 {\displaystyle \Psi_2 
=-{\xi_{,u}\eta_{,v}\over2f(\xi-\eta)}
\Big(2(\xi-\eta)\psi_{,\xi}\,\psi_{,\eta} +\psi_{,\xi} -\psi_{,\eta}\Big)}
\\[10pt]
 {\displaystyle \Psi_4 =-\xi_{,u}\,\psi_{,\xi}\delta(u)
-{{\xi_{,u}^2}\over f}\left(\psi_{,\xi\xi}
-2(\xi-\eta)\,{\psi_{,\xi}}^3 +3{\psi_{,\xi}}^2\right).} 
 \end{array}
 \label{WeylCompts}
 \end{equation}

\section{The singularity in the interaction region}

Now consider the singularity which occurs in region~IV on the spacelike
hypersurface on which \ $\alpha={1\over2}(\xi-\eta)=0$. \ It can be shown (see
references in \cite{Griff91}) that this is generically a scalar polynomial
curvature singularity on which the invariant \
$\Psi_0\Psi_4-4\Psi_1\Psi_3+3\Psi_2^2$ \ diverges. However, families of solutions
for colliding plane waves exist, even having infinitely many parameters, in which
this invariant does not diverge. In these cases, the singularity at $\alpha=0$
corresponds to a Killing--Cauchy horizon through which the space-time can be
extended.

The general solution of the main equation (\ref{EPD}) can be expressed (exactly
as for vacuum Gowdy cosmologies) in terms of the coordinates \
$\alpha={1\over2}(\xi-\eta)$ \ and \ $\beta={1\over2}(\xi+\eta)$ \ in the form 
 \begin{equation}
 \psi=\int_{-\infty}^\infty
\Big[a(\omega)J_0(\omega\alpha)+b(\omega)Y_0(\omega\alpha)\Big]
e^{i\omega\beta}\d\omega ,
 \label{Gowdy}
 \end{equation}
 where $J_0(\omega\alpha)$ and $Y_0(\omega\alpha)$ are Bessel functions of the
first and second kinds of zero order, and $a(\omega)$ and $b(\omega)$ are
arbitrary functions of the parameter~$\omega$. The terms which involve
$J_0(\omega\alpha)$ remain well behaved on the singularity at \ $\alpha=0$, \
while the terms which involve $Y_0(\omega\alpha)$ diverge logarithmically (the
presence of some terms of this type is necessary to satisfy the initial data).

In his analysis of the asymptotic behaviour of colliding plane wave space-times,
Yurtsever \cite{Yurts88c} has shown that, for any timelike geodesic which
approaches the singularity $\alpha=0$, the metric asymptotically approaches that
of a particular Kasner solution. However, the particular Kasner exponents vary
for geodesics which approach different points on the singularity. It follows that
the focussing singularity is generally a curvature singularity. The only
exception is that in which the metric approaches that of the degenerate (flat)
Kasner solution at all points over the hypersurface $\alpha=0$. In this case, the
focussing hypersurface is a Killing--Cauchy horizon across which space-time can
be extended. Such solutions are characterised by the asymptotic behaviour  
 \begin{equation}
 \label{HorizonCondition}
\psi(\xi,\eta)\simeq \log\alpha-B(\beta) \qquad \hbox{or} \qquad 
\psi(\xi,\eta)\simeq B(\beta)
 \label{asbeh}
 \end{equation}
 as $\alpha\to 0$, where $B(\beta)$ is an arbitrary function which is independent
of $\alpha$. However, the substitution \ $\psi\to\log\alpha-\psi$ \ simply
corresponds to an interchange of the $x,y$ coordinates. Thus, the two possible
conditions in (\ref{HorizonCondition}) are physically equivalent, and the second
can be adopted as the criteria for a colliding plane wave solution with a
horizon.

In considering perturbations of such solutions, Yurtsever \cite{Yurts88c} has
rigorously shown that arbitrarily small changes in the amplitude functions
(specifically in the coefficients $b(\omega)$ in (\ref{Gowdy})) will cause
variations in the Kasner parameters. He has reasonably concluded that, although
there are colliding plane wave space-times which contain a Killing--Cauchy
horizon rather than a space-like curvature singularity, these space-times are
unstable against small perturbations of the initial data and that `generic'
initial data leads to curvature singularities. However, it may be observed that
the perturbations which induce a curvature singularity in the interaction region
also generally have the effect that the relevant Weyl tensor component becomes
unbounded at the fold singularities in the initial regions II and III. In other
words, the components which introduce a curvature singularity into a space-time
which otherwise has a horizon are associated with unphysical initial data. Thus,
the question that still needs to be addressed is that of the stability of the
horizon with respect to perturbations which correspond to bounded variations of
the initial gravitational waves prior to their collision.

\section{Explicit solutions with horizons}

Now consider the colliding plane wave solutions \cite{FerIba87} for which 
 \begin{equation}
 \psi={(1+a)\over2}\log\alpha 
-{k_{\scriptscriptstyle+}\over2} \cosh^{-1}\left({1-\beta\over{\alpha}}\right) 
-{k_{\scriptscriptstyle-}\over2} \cosh^{-1}\left({1+\beta\over{\alpha}}\right) ,
 \label{CPWcompts}
 \end{equation} 
 where $a$, $k_{\scriptscriptstyle+}$ and $k_{\scriptscriptstyle-}$ are
constants. When \ $a=0$ \ and \
$k_{\scriptscriptstyle+}=k_{\scriptscriptstyle-}=1$, \ this is the Khan--Penrose
solution \cite{KhaPen71} which represents the collision of purely impulsive
waves. It also includes the degenerate Ferrari--Ib\'a\~nez solutions
\cite{FerIba88}, for which the interaction region is locally isomorphic to part
of the Schwarzschild solution inside the horizon.

By showing that, as \ $\alpha\to0$, \ the scalar invariant behaves as 
 $$ \Psi_0\Psi_4+3\Psi_2^2 \to{{\xi_{,u}}^2{\eta_{,v}}^2\over2^8f^2} 
\Big((a+k_{\scriptscriptstyle+}+k_{\scriptscriptstyle-})^2-1\Big)^2
\Big((a+k_{\scriptscriptstyle+}+k_{\scriptscriptstyle-})^2+3\Big) \alpha^{-4}, $$ 
 Feinstein and Ib\'a\~nez \cite{FeiIba89} have shown that colliding plane wave
solutions given by (\ref{CPWcompts}) have a horizon at \ $\alpha=0$ \ in
region~IV rather than a curvature singularity if \
$a+k_{\scriptscriptstyle+}+k_{\scriptscriptstyle-}=\pm1$. \ (This also
follows from the condition (\ref{HorizonCondition}).)

It can be seen that the last two terms in the expansion (\ref{CPWcompts}) are
required to describe the necessary behaviour of the solution near the wavefronts
$\xi=1$ and $\eta=-1$ respectively, on which they also become zero. Also, all
three components of (\ref{CPWcompts}) diverge logarithmically as $\alpha\to0$.
Thus, these terms must include components which arise from the
$Y_0(\omega\alpha)$ terms in (\ref{Gowdy}) together with possible components
which arise from the regular terms. In addition, it is clear that the solutions
(\ref{CPWcompts}) arise from the initial data functions 
 \begin{equation}
 \begin{array}{l}
 \psi_{\scriptscriptstyle+}(\xi) ={(1+a)\over2}\log\left({1+\xi\over2}\right) 
-{k_{\scriptscriptstyle+}\over2} \cosh^{-1}\left({3-\xi\over1+\xi}\right), 
\\[12pt]
 \psi_{\scriptscriptstyle-}(\eta)
={(1+a)\over2}\log\left({1-\eta\over2}\right)  -{k_{\scriptscriptstyle-}\over2}
\cosh^{-1}\left({3+\eta\over1-\eta}\right). 
 \end{array}
 \label{initialFIdata}
 \end{equation} 
 Thus the constants are determined by the initial data. Moreover, to satisfy the
initial conditions for colliding plane waves, it can be shown that
$k_{\scriptscriptstyle\pm}$ must satisfy \
$1\le|k_{\scriptscriptstyle\pm}|<\sqrt2$. \ It can then be seen that the initial
wave in region~II is 
 $$ \Psi_{4\scriptscriptstyle+} = 
-{k_{\scriptscriptstyle+}\xi_{,u}\over2\sqrt{2(1-\xi)}}\,\delta(u)
 -{{\xi_{,u}}^2\over4(1+\xi)^2} \left[ a(1-a^2)
-{3\sqrt2a^2k_{\scriptscriptstyle+}\over\sqrt{1-\xi}}
-{6\,a\,k_{\scriptscriptstyle+}^2\over(1-\xi)}
+{2\sqrt2k_{\scriptscriptstyle+}(1-k_{\scriptscriptstyle+}^2)
\over(1-\xi)^{3/2}} \right], $$ 
 which, as \ $\xi\to-1$, \ behaves as 
 $$ \Psi_{4\scriptscriptstyle+}\to
 {(a+k_{\scriptscriptstyle+})\over4}
\left[(a+k_{\scriptscriptstyle+})^2-1\right]  {{\xi_{,u}}^2\over(1+\xi)^2}. $$ 
 This clearly diverges unless $a+k_{\scriptscriptstyle+}$ has the values 0, +1
or $-1$. The opposing wave similarly diverges as $\eta\to1$ unless
$a+k_{\scriptscriptstyle-}$ has the values 0, +1 or $-1$. Thus, within this
family of solutions, initial waves which are bounded as $\xi\to-1$ or $\eta\to1$
and which give rise to solutions with a Killing--Cauchy horizon can only occur in
a very limited number of possible cases. In fact, the only possibility up to a
sign occurs when \ $a=-1$ \ and \ $k_{\scriptscriptstyle+}=
k_{\scriptscriptstyle-}=1$. \ This is precisely the degenerate
Ferrari--Ib\'a\~nez solution with a horizon~\cite{FerIba88}.

Feinstein and Ib\'a\~nez \cite{FeiIba89} have also shown that the above result
can be extended to solutions in which $\psi$ is given by (\ref{CPWcompts}) plus
an arbitrary regular function that can be expressed as \
$\int_{-\infty}^\infty a(\omega)J_0(\omega\alpha) e^{i\omega\beta}\d\omega$ \ as
in (\ref{Gowdy}). (This is currently the widest explicitly known class of
colliding plane wave solutions with a horizon.) It follows that perturbations
which correspond to variations in the arbitrary function $a(\omega)$ do not alter
the character of the horizon at \ \hbox{$\alpha=0$}. \ On the other hand,
perturbations of any of the parameters $a$ or $k_{\scriptscriptstyle\pm}$ lead
to the occurrence of a curvature singularity in region~IV and also give rise to
to divergencies in the initial waves. Perturbations corresponding to additional
components of $b(\omega)$ in (\ref{Gowdy}) similarly introduce both a curvature
singularity in the interaction region and divergencies in the initial waves.

\section{The stability of solutions with horizons}

From the above discussion, it may be concluded that some perturbations which
indicate that horizons in colliding plane wave space-times are unstable
correspond to unbounded perturbations of the initial data. On the other hand, such
horizons appear to be stable with respect to perturbations which correspond to
arbitrary variations in the function $a(\omega)$. In this case, however, although
the corresponding perturbations of the initial waves are bounded, this
corresponds to a situation in which the perturbation of one wave is exactly
correlated with the perturbation of the other. Thus, it is not possible to
perturb the initial data on one characteristic while keeping constant those on
the other and at the same time preserving the character of the horizon in the
interaction region. It follows that, for the above explicit solutions at least,
the horizon must be {\em unstable} with respect to small {\em independent}
perturbations of the initial data.

To demonstrate the generic instability of any colliding plane wave solution with a
horizon, consider a perturbation for which the initial data is given by 
 \begin{equation}
 \psi_{\scriptscriptstyle+}(\xi) = \psi_{b\scriptscriptstyle+}(\xi)
+R_{\scriptscriptstyle+}(\xi),  \qquad
 \psi_{\scriptscriptstyle-}(\eta) = \psi_{b\scriptscriptstyle-}(\eta),
 \label{perturbation}
 \end{equation}
 where $\psi_{b\scriptscriptstyle+}(\xi)$ and $\psi_{b\scriptscriptstyle-}(\eta)$
are the initial data which lead to a background solution with a horizon, and 
$R_{\scriptscriptstyle+}(\xi)$ is a perturbation function which is bounded
for \ $\xi\in[-1,1]$ \ and such that \ $R_{{\scriptscriptstyle+},\xi}\to0$ \ as \
$\xi\to-1$. i.e. a perturbation has been included in one initial wave only and,
when \ $R_{\scriptscriptstyle+}(\xi)=0$, \ a horizon occurs in the interaction
region.

Use can now be made of the solution of the characteristic initial value problem
for colliding plane waves given by Hauser and Ernst \cite{HauErn89a} and
implemented in \cite{GriSan02}. However, it is appropriate to present this in a
slightly modified form which is more clearly related to that of the general
non-linear case described in \cite{AleGri04}. In this method, it is first
necessary to calculate the initial spectral amplitude functions
$A_{\scriptscriptstyle+}(\zeta_{\scriptscriptstyle+})$ and
$A_{\scriptscriptstyle-}(\zeta_{\scriptscriptstyle-})$ from the initial data
functions $\psi_{\scriptscriptstyle+}(\xi)$ and
$\psi_{\scriptscriptstyle-}(\eta)$ using the integrals 
 \begin{equation}
 A_{\scriptscriptstyle+}(\zeta_{\scriptscriptstyle+})=
-{1\over\pi\sqrt{1-\zeta_{\scriptscriptstyle+}}}
\int_{\zeta_{\scriptscriptstyle+}}^1 {\psi_{{\scriptscriptstyle+},\xi}
\over\sqrt{\xi-\zeta_{\scriptscriptstyle+}}} \,\d\xi,
\qquad
 A_{\scriptscriptstyle-}(\zeta_{\scriptscriptstyle-})=
{1\over\pi\sqrt{\zeta_{\scriptscriptstyle-}+1}}
\int_{-1}^{\zeta_{\scriptscriptstyle-}} {\psi_{{\scriptscriptstyle-},\eta}
\over\sqrt{\zeta_{\scriptscriptstyle-}-\eta}} \,\d\eta. 
 \label{Abel2}
 \end{equation} 
 The solution in the interaction region can then be expressed as 
 \begin{equation}
 \psi(\xi,\eta)=\int_\xi^1 
{\sqrt{1-\zeta_{\scriptscriptstyle+}} \sqrt{\zeta_{\scriptscriptstyle+}+1}
\over\sqrt{\zeta_{\scriptscriptstyle+}-\xi}
\sqrt{\zeta_{\scriptscriptstyle+}-\eta}} 
\,A_{\scriptscriptstyle+}(\zeta_{\scriptscriptstyle+})
\,\d\zeta_{\scriptscriptstyle+}
 +\int_{-1}^\eta {\sqrt{1-\zeta_{\scriptscriptstyle-}}
\sqrt{\zeta_{\scriptscriptstyle-}+1}
\over\sqrt{\xi-\zeta_{\scriptscriptstyle-}}
\sqrt{\eta-\zeta_{\scriptscriptstyle-}}}
\,A_{\scriptscriptstyle-}(\zeta_{\scriptscriptstyle-})
\,\d\zeta_{\scriptscriptstyle-}. 
 \label{gensolution}
 \end{equation}

This method, which is based on the Abel transform, is linear. Thus, when the
perturbation vanishes (i.e. when $R_{\scriptscriptstyle+}(\xi)=0$), the resulting
solution would be just the background solution $\psi_b(\xi,\eta)$ which contains
a horizon. However, the perturbation term in (\ref{perturbation}) would lead to
an additional perturbation component in the initial spectral data given by 
 $$ A_{p\scriptscriptstyle+}(\zeta_{\scriptscriptstyle+})=
-{1\over\pi\sqrt{1-\zeta_{\scriptscriptstyle+}}}
\int_{\zeta_{\scriptscriptstyle+}}^1 {R_{{\scriptscriptstyle+},\xi}
\over\sqrt{\xi-\zeta_{\scriptscriptstyle+}}} \,\d\xi,
\qquad A_{p\scriptscriptstyle-}(\zeta_{\scriptscriptstyle-})=0, $$ 
 and this would lead to the solution in the interaction region being given by 
 \begin{equation}
 \psi=\psi_b+\psi_p, \qquad \hbox{where} \qquad
 \psi_p= \int_\xi^1 
{\sqrt{1-\zeta_{\scriptscriptstyle+}} \sqrt{\zeta_{\scriptscriptstyle+}+1}
\over\sqrt{\zeta_{\scriptscriptstyle+}-\xi}
\sqrt{\zeta_{\scriptscriptstyle+}-\eta}} 
\,A_{p\scriptscriptstyle+}(\zeta_{\scriptscriptstyle+})
\,\d\zeta_{\scriptscriptstyle+}. 
 \label{pertsoln}
 \end{equation} 
 It can then be shown that the scalar invariant \ $\Psi_0\Psi_4+3\Psi_2^2$ \
diverges as $\alpha\to0$ (actually, it is the component $\Psi_2$ which diverges),
unless \ $\psi_{p,\xi}-\psi_{p,\eta}\to0$ \ in this limit. (This is equivalent to
the condition (\ref{asbeh}).)

However, the perturbation $R_{\scriptscriptstyle+}(\xi)$ is arbitrary and it is
therefore possible to consider the component
$A_{p\scriptscriptstyle+}(\zeta_{\scriptscriptstyle+})$ to be an arbitrary
bounded function satisfying \ $\lim_{\zeta_+\to1} \big(
\sqrt{1-\zeta_{\scriptscriptstyle+}}\,
A_{p\scriptscriptstyle+}(\zeta_{\scriptscriptstyle+}) \big) =0$. \ Moreover, it
can be seen that 
 \begin{equation}
 \psi_{p,\xi}-\psi_{p,\eta} = \int_\xi^1
{\sqrt{\zeta_{\scriptscriptstyle+}-\eta}\over
\sqrt{\zeta_{\scriptscriptstyle+}-\xi}} 
\,{\d\over\d\zeta_{\scriptscriptstyle+}}\left[
{\sqrt{1-\zeta_{\scriptscriptstyle+}} \sqrt{\zeta_{\scriptscriptstyle+}+1}
\over(\zeta_{\scriptscriptstyle+}-\eta)} 
\,A_{p\scriptscriptstyle+}(\zeta_{\scriptscriptstyle+})
\right]\,\d\zeta_{\scriptscriptstyle+} , 
 \label{pertexpn}
 \end{equation} 
 which does not necessarily approach zero as \ $\alpha\to0$ \ for an arbitrary
choice of $A_{p\scriptscriptstyle+}(\zeta_{\scriptscriptstyle+})$. (Indeed, it
may well be unbounded, but this stronger result is not needed.) It can therefore
be concluded that a perturbation of one initial wave will generically lead to a
scalar polynomial curvature singularity in the interaction region in place of the
horizon. i.e. the Killing--Cauchy horizon is {\em unstable} with respect to the
perturbation~(\ref{perturbation}).

\section{Conclusion}

Colliding plane wave solutions which contain a Killing--Cauchy horizon, only
occur in very particular circumstances. Moreover, colliding plane wave
space-times which are generated by bounded initial waves (apart from impulsive
components on the shock fronts) and possess a Killing--Cauchy horizon form a
highly restricted subclass of these solutions. It has been shown above that the
only permitted perturbations of this subclass which preserve such a horizon are
those in which perturbations of the two approaching waves are strongly correlated.
However, when the initial waves are perturbed independently, then the horizon in
the interaction region is necessarily replaced by a scalar polynomial curvature
singularity. It is therefore concluded that the Killing--Cauchy horizon which
sometimes occurs in colliding plane wave space-times is unstable with respect to
general bounded perturbations of the initial waves.

The above discussion has concentrated entirely on the ``linear case'' in which
the approaching gravitational waves have constant and aligned polarization.
However, it is expected that the behaviour of the nonlinear case in which the
initial waves have variable or non-aligned polarizations, or include
electromagnetic waves, would be qualitatively similar. This expectation is
supported by the arguments of Yurtsever~\cite{Yurts89} for the vacuum case, but
further analysis is required.

\section*{Acknowledgments}

It is a pleasure to thank Professor G. A. Alekseev for many discussions and a
lengthy correspondence on this topic.

\end{document}